\documentclass[twocolumn]{aastex62}

\usepackage{color}
\usepackage{url}

\submitjournal{ApJL}

\shorttitle{A Jet Signature}
\shortauthors{{Mooley} et al.}

\begin{document}

% I would not use the word classic, but "Strong" would work here. Or nothing at all.
\title{A Strong Jet Signature in the Late-Time Lightcurve of GW170817}

\correspondingauthor{K. P. Mooley}
\email{kunal@astro.caltech.edu}

% \author{Authors}
% \affil{Affiliations}

\author[0000-0002-2557-5180]{K. P. Mooley}
\altaffiliation{Jansky Fellow (NRAO/Caltech).}
\affil{National Radio Astronomy Observatory, Socorro, New Mexico 87801, USA}
\affil{Caltech, 1200 E. California Blvd. MC 249-17, Pasadena, CA 91125, USA}

\author{D. A. Frail}
\affil{National Radio Astronomy Observatory, Socorro, New Mexico 87801, USA}

\author[0000-0003-0699-7019]{D. Dobie}
\affil{Sydney Institute for Astronomy, School of Physics, University of Sydney, Sydney, New South Wales 2006, Australia.}
% \affil{ARC Centre of Excellence for All-sky Astrophysics (CAASTRO), Australia.}
\affil{ATNF, CSIRO Astronomy and Space Science, PO Box 76, Epping, New South Wales 1710, Australia}

\author[0000-0002-9994-1593]{E. Lenc}
\affil{ATNF, CSIRO Astronomy and Space Science, PO Box 76, Epping, New South Wales 1710, Australia}

\author[0000-0001-8104-3536]{A. Corsi}
\affil{Department of Physics and Astronomy, Texas Tech University, Box 41051, Lubbock, TX 79409-1051, USA}

\author{K. De}
\affil{Caltech, 1200 E. California Blvd. MC 249-17, Pasadena, CA 91125, USA}

\author{A.J. Nayana}
\affil{National Centre for Radio Astrophysics, Tata Institute of Fundamental Research, Pune University Campus, Ganeshkhind Pune 411007, India}

\author{S. Makhathini}
\affil{South African Radio Astronomy Observatory (SARAO), Pinelands 7405, South Africa}
\affil{Department of Physics \& Electronics, Rhodes University, Grahamstown, 6139, South Africa}

\author{I. Heywood}
\affil{Denys Wilkinson Building, Keble Road, Oxford OX1 3RH, UK}
\affil{Department of Physics \& Electronics, Rhodes University, Grahamstown, 6139, South Africa}

\author[0000-0002-2686-438X]{T. Murphy}
\affil{Sydney Institute for Astronomy, School of Physics, University of Sydney, Sydney, New South Wales 2006, Australia.}
% \affil{ARC Centre of Excellence for All-sky Astrophysics (CAASTRO), Australia.}

\author[0000-0001-6295-2881]{D. L. Kaplan}
\affil{Department of Physics, University of Wisconsin - Milwaukee, Milwaukee, Wisconsin 53201, USA.}

\author{P. Chandra}
\affil{National Centre for Radio Astrophysics, Tata Institute of Fundamental Research, Pune University Campus, Ganeshkhind Pune 411007, India}

\author{O. Smirnov}
\affil{Department of Physics \& Electronics, Rhodes University, Grahamstown, 6139, South Africa}
\affil{South African Radio Astronomy Observatory (SARAO), Pinelands 7405, South Africa}

% \author[0000-0002-0494-192X]{C. Lynch}
% \affil{Sydney Institute for Astronomy, School of Physics, University of Sydney, Sydney, New South Wales 2006, Australia.}
% \affil{ARC Centre of Excellence for All-sky Astrophysics (CAASTRO), Australia.}

\author{E. Nakar}
\affil{The Raymond and Beverly Sackler School of Physics and Astronomy, Tel Aviv University, Tel Aviv 69978, Israel}

\author[0000-0002-7083-4049]{G. Hallinan}
\affil{Caltech, 1200 E. California Blvd. MC 249-17, Pasadena, CA 91125, USA}

\author{F. Camilo}
\affil{South African Radio Astronomy Observatory (SARAO), Pinelands 7405, South Africa}

\author{R. Fender}
\affil{Denys Wilkinson Building, Keble Road, Oxford OX1 3RH, UK}
\affil{Department of Astronomy, University of Cape Town, Private Bag X3, Rondebosch 7701, South Africa}

\author{S. Goedhart}
\affil{South African Radio Astronomy Observatory (SARAO), Pinelands 7405, South Africa}

\author{P. Groot}
\affil{Department of Astrophysics/IMAPP, Radboud University Nijmegen, P. O. Box 9010, 6500 GL Nijmegen, The Netherlands}
\affil{Department of Astronomy, University of Cape Town, Private Bag X3, Rondebosch 7701, South Africa}

% \author{V. Bhalerao}
% \affil{Department of Physics, Indian Institute of Technology Bombay, Mumbai 400076,
% India}

\author[0000-0002-5619-4938]{M. M. Kasliwal}
\affil{Caltech, 1200 E. California Blvd. MC 249-17, Pasadena, CA 91125, USA}

\author[0000-0001-5390-8563]{S. R. Kulkarni}
\affil{Caltech, 1200 E. California Blvd. MC 249-17, Pasadena, CA 91125, USA}

\author[0000-0002-6896-1655]{P.A. Woudt}
\affil{Department of Astronomy, University of Cape Town, Private Bag X3, Rondebosch 7701, South Africa}

\begin{abstract}
%The gradual rise in the radio and X-ray light curves of the binary neutron star merger GW170817, seen up to 150 days post-merger, gave evidence for a wide-angle mildly relativistic outflow or cocoon-dominated emission. Such an outflow could be generated by the interaction of a jet with the neutron-rich material dynamically ejected during the merger, but the light curve data reported to date have not given a clear insight into the fate of the putative jet (whether it successfully escaped the dynamical ejecta or got choked). Here we report deep radio observations of GW170817 with the VLA and ATCA out to almost 300 days post-merger that we use to characterize the decay phase of the light curve. We are thus able to probe the geometry and dynamical state of the outflow, as has been done two decades ago for the late-time light curves of gamma-ray burst (GRB) afterglows. We find that the decay of the light curve is consistent with $t^{-p}$, where $p$ is the electron power law index, and the transition between the power-law rise an decay is relatively sharp (occurs over a span of $<$80 days). This provides strong evidence that the late-time afterglow is dominated by a narrow and energetic jet, and strengthens the link between binary neutron star mergers and short-hard GRBs. Our results are consistent with those from our companion VLBI paper reporting superluminal motion in GW170817.

% Revised abstract that is more in a style of an observational paper - Dale

We present new 0.6--10~GHz observations of the binary neutron star merger \object{GW170817} covering the period up to 300 days post-merger, taken with the upgraded Karl G. Jansky Very Large Array, the Australia Telescope Compact Array, the Giant Metrewave Radio Telescope and the MeerKAT telescope. We use these data to precisely characterize the decay phase of the late-time radio light curve. We find that the temporal decay is consistent with a power-law slope of $t^{-2.2}$, and that the transition between the power-law rise and decay is relatively sharp.
% (occurring over a span of $<$80 days). 
Such a slope cannot be produced by a quasi-isotropic (cocoon-dominated) outflow, but is instead the classic signature of a relativistic jet. This provides strong observational evidence that \object{GW170817} produced a successful jet, and directly demonstrates the link between binary neutron star mergers and short-hard GRBs. Using simple analytical arguments, we derive constraints on the geometry and the jet opening angle of \object{GW170817}. These results are consistent with those from our companion Very Long Baseline Interferometry (VLBI) paper, reporting superluminal motion in \object{GW170817}.

%We present 2--9~GHz radio observations of \object{GW170817} covering the period 125--200 days post-merger, taken with the Australia Telescope Compact Array and the Karl G. Jansky Very Large Array. Our observations demonstrate that the radio afterglow peaked at $149\pm2$ days post-merger and is now declining in flux density.  We see no evidence for evolution in the radio-only spectral index, which remains consistent with optically-thin synchrotron emission connecting the radio, optical, and X-ray regimes.   The peak  implies a total energy in the synchrotron-emitting component of a ${\rm few}\times 10^{50}$\,erg. The temporal decay rate is most consistent with mildly- or non-relativistic material and we do not see evidence for a very energetic off-axis jet, but we cannot distinguish between a lower-energy jet and more isotropic emission. 
\end{abstract}


\section{Introduction} \label{sec:intro}

The detection of gravitational waves (GW) from the binary neutron star merger event \object{GW170817} \citep{2017PhRvL.119p1101A}, was accompanied by two distinct electromagnetic (EM) counterparts \citep{{2017ApJ...848L..12A}}. The first EM counterpart was a short-lived, thermal-like component. It initially had bright optical/UV emission that faded on a timescale of a few days \citep{2017Sci...358.1556C, Evans1565,2017Natur.551...64A,2017ApJ...850L...1L,2017ApJ...848L..16S,2017ApJ...848L..24V}, to be replaced by redder emission which dominated the bolometric luminosity until it too faded on a timescale of a few weeks \citep{2017ApJ...848L..27T,2017ApJ...848L..17C,2017Sci...358.1570D,2017Sci...358.1559K}. Optical/near-IR spectroscopy detected the spectral fingerprints of r-process elements \citep{Pian2017,smartt2017}, 
% mccully20017,chornock2017,kasliwal2017,shappee2017,buckley2017
and a broad consensus has formed that this thermal component was a ``kilonova'' powered by the dynamic ejecta from the merger event \citep[e.g.][]{2017Natur.551...80K}.

% The second EM counterpart was initially 
Additionally, the EM counterpart was also detected as a prompt, low-luminosity burst of gamma-rays with duration $\sim$2 s but delayed from the GW merger event by 1.7 s \citep{2017ApJ...848L..14G}. This was followed by the discovery of non-thermal ``afterglow'' emission at X-ray \citep{2017ApJ...848L..25H, 2017ApJ...848L..20M,2017Natur.551...71T} and radio wavelengths \citep{2017Sci...358.1579H}, substantially delayed by 9 and 16 days, respectively. Both the prompt and the afterglow emission are thought to be generated in a relativistic shock, but a consensus on the nature of this second EM component has been slower to emerge. Two viable models were eliminated based on the early data. From the delayed onset of the afterglow it was conclusively shown that \object{GW170817} was not a classical short-hard gamma-ray burst (SHB) with an on-axis jet \citep[e.g.][]{2017Sci...358.1559K}. Furthermore, an off-axis jet with a uniform or ``top-hat'' geometry was ruled out by the slow rise ($t^{+0.8}$) of the radio emission up to 100 days post-merger \citep{2018Natur.554..207M}, later confirmed at both optical and X-ray wavelengths \citep{2018ApJ...856L..18M, 2018MNRAS.478L..18T, 2018arXiv180102669L,ruan2018}.

There still remain two competing models for the prompt and afterglow emission of \object{GW170817}, both of which are well motivated physically. Both scenarios launch an ultra-relativistic jet (bulk Lorentz factor $\Gamma\sim$100), pointing away from the Earth, that interacts with the neutron-rich material dynamically ejected during the merger to give rise to a mildly relativistic ($\Gamma\sim$4) outflow (a.k.a ``cocoon'') moving in the direction of the Earth.
The mildly relativistic material is likely responsible for the gamma-ray signal \cite[e.g.][]{2017Sci...358.1559K,2018MNRAS.473..576G,matsumoto2018} and is primarly responsible for the slow rise ($t^{+0.8}$) of the afterglow at early times \citep[e.g.][]{2018Natur.554..207M}. 
However, in one scenario the jet successfully escapes the dynamical ejecta, while in the other it fails to do so.
In the literature the former case has been referred to as the successful jet-cocoon model or the structured jet model (where the successful jet is considered to be a narrow core with a ``sheath'' of lower Lorentz factor material), in which the jet dominates the late-time afterglow, whereas the latter case is that of a choked-jet, in which the afterglow is cocoon-dominated at all times.
In the successful jet scenario, an observer located along the axis of the jet likely sees a regular SHB, while in the choked jet scenario they do not.

%no observer sees a regular SHB.
In the discussion that follows on the late-time light curves, we will refer to these as jet-dominated and cocoon-dominated outflows. 
Both outflow models require significant azimuthal and radial structure to explain the rise of the afterglow light curve \citep{nakar-piran2018,2018arXiv180409345X}, but the open question that we hope to address here is whether the relativistic jet survived.
%both outflow models require significant azimuthal or radial structure to explain the rise of the afterglow light curve \citep{2018arXiv180409345X}. The jet is likely composed of a narrow core with a ``sheath'' of lower Lortenz factor material with a larger opening angle, while the cocoon requires ejecta with a stratified (power-law) distribution of velocities.

Several experimental tests have been suggested to distinguish between these two alternative scenarios \citep{2018MNRAS.tmp.1159G,2018arXiv180307595N,2018PhRvL.120x1103L}. Elsewhere we report on \object{GW170817} polarization measurements and our high angular resolution imaging \citep{2018ApJ...861L..10C,2018arXiv180609693M}. In this  paper we focus on the continuum intensity of the afterglow light curve at late times. Several authors have noted that the rising portion of the light curve has limited discriminating power since the lower Lorentz factor ejecta dominate the emission in both models
\citep{2018arXiv180307595N, 2018ApJ...856L..18M,davanzo2018,2018MNRAS.tmp.1159G}. However, as first noted by \cite{2018ApJ...858L..15D}, and subsequently confirmed by \cite{2018arXiv180502870A}, the afterglow light curve peaked
% at all wavelengths 
around day 150 post-merger and has begun to decline. Our motivation for this work has been to characterize this decay phase (as attempted by previous studies that reported the peak and decline) and to see whether the geometry and dynamical state of the outflow can be inferred as it has been done in the past with the late-time light curves of long-duration gamma-ray burst (GRB) afterglows \citep{1998ApJ...499..301M,1999ApJ...519L..17S,2000ApJ...538..187L}. 

In this paper we present further radio observations of \object{GW170817} using the NSF's Karl G.\ Jansky Very Large Array (VLA), the Australia Telescope Compact Array (ATCA), the upgraded Giant Metrewave Radio Telescope (uGMRT) and new observations from the MeerKAT telescope, covering the period 180--300 days post-merger. With this longer time-baseline we are able to accurately constrain the late-time power-law decay index and compare the results against expectations for a cocoon-dominated versus a jet-dominated outflow. In \S\ref{sec:data} we describe the observations and data reduction techniques employed, \S\ref{sec:results} gives the spectral and light curve analysis with the interpretation based on widely used theory of GRB afterglows given in \S\ref{sec:discussion}, and we end with the summary and conclusions in \S\ref{sec:conclusion}.

\begin{deluxetable}{cccccccC}[b!]
\tablecaption{Radio observations of \object{GW170817} during the light curve decline\label{tab:radiodata}}
\tablecolumns{6}
\tablenum{1}
\tablewidth{0pt}
\tablehead{
\colhead{UT date} & \colhead{$\Delta$T$^{\dag}$} & \colhead{Telescope} & \colhead{$\nu$} & \colhead{$F_\nu$} & \colhead{$\sigma_\nu$}\\
\colhead{} & \colhead{(d)} & \colhead{} &\colhead{(GHz)} & \colhead{($\mu$Jy)} & \colhead{($\mu$Jy)}
}
\startdata
2018 Feb 16            & 183    & uGMRT   & 0.65 & 211  & 34  \\
2018 Mar 02            & 197    & MeerKAT & 1.3  & 160  & 20  \\
2018 Mar 21            & 216    & VLA     & 10   & 36.3 & 3.6 \\
2018 Mar 25--26        & 220    & VLA     & 3    & 64.7 & 2.7 \\
2018 Mar 27            & 222    & ATCA$^\ddag$   & 7.25 & 39.7 & 7.2 \\
2018 Apr 26--May 05    & 257    & MeerKAT & 1.3  & 65.8 & 7.2 \\
2018 May 11--12        & 267    & VLA     & 3    & 40.3 & 2.7 \\
2018 May 11            & 267    & ATCA$^\ddag$    & 7.25 & 25.0 & 4.1 \\
2018 May 13-25         & 275    & uGMRT   & 0.65 & $<$153 & ... \\
2018 Jun 07            & 294    & VLA     & 3    & 31.2 & 3.6 \\
2018 Jun 11            & 298    & ATCA$^\ddag$    & 7.25 & 23.4 & 4.2 \\
% 2018 Jun 19--30        & 310    & uGMRT   & 0.65 & ...  & ... \\
\enddata
\tablenotetext{\dag}{Mean epoch, days post-merger.}
\tablenotetext{\ddag}{The ATCA flux densities of GW170817 have a correction factor of 1.25 applied (i.e. the values have been decreased by 25\%; see \S\ref{subsec:spectral} for details).}
% \tablenotetext{a}{With the 6C configuration (maximum baselines of 6\,km) and program CX391 (PI: T.~Murphy).}
% \tablenotetext{b}{With the 750A configuration (maximum baseline of 3.75\,km) and program CX394 (PI: E.~Troja).}
% \tablenotetext{c}{Insufficient data quality}
% \tablenotetext{d}{With the 750B configuration (maximum baseline of 4.5\,km) and program CX394 (PI: E.~Troja).}
% \tablenotetext{e}{With the A configuration (maximum baseline of 27\,km) under a Director Discretionary Time program (VLA/17B-397;
% PI: K. Mooley).}
\end{deluxetable}

\section{Observations and Data reduction} \label{sec:data}

Our observations of GW170817 carried out over the decline of the light curve are described below, and the observing log together with the flux densities of GW170817 are reported in Table~\ref{tab:radiodata}.
While our new VLA and ATCA observations span frequencies between 2--12 GHz, the addition of MeerKAT and uGMRT data gives us important spectral coverage below 2 GHz, which not only probes the low-frequency behavior of GW170817, but also gives very precise measurements of the radio spectral indices.

\subsection{VLA} \label{subsec:vla}
We observed \object{GW170817} on 2018 Mar 21, Mar 25--26, May 11--12, and Jun 07 with the VLA (PI: Corsi). 
% The observations are summarized in Table~\ref{tab:radiodata}. 
The Wideband Interferometric Digital Architecture (WIDAR) correlator was used at S band (2--4\,GHz) and X band (8--12\,GHz) to maximize sensitivity. We used \object[PKS J1248-1959]{PKS~J1248$-$1959} as the phase calibrator and \object{3C286} as the flux density and bandpass calibrator. The data were calibrated and flagged for radio frequency interference (RFI) using the NRAO {CASA} \citep{2007ASPC..376..127M} pipeline. We then split and imaged the target data using the CASA tasks {\tt split} and {\tt clean}.
% The VLA flux densities of GW170817 are reported in Table~\ref{tab:radiodata}.

\subsection{ATCA} \label{subsec:atca}

We observed \object{GW170817} with the ATCA (PI: Dobie, Troja) at three epochs between 2018 Mar to 2018 Jun. 
% A log of observations is given in Table~\ref{tab:radiodata}. 
We determined the flux scale and bandpass response for all epochs using the ATCA primary calibrator \object[PKS B1934-638]{PKS~B1934$-$638}. Observations of \object[PKS B1245-197]{PKS~B1245$-$197} were used to calibrate the complex gains. All observations used two bands of 2048 MHz centered at 5.5 and 9.0\,GHz. 

We reduced the visibility data using standard MIRIAD \citep{1995ASPC...77..433S} routines. 
% The calibrated visibility data were split into the 5.5 and 9.0\,GHz bands, averaged to 32\,MHz channels, and imported into DIFMAP \citep{1997ASPC..125...77S}. 
The calibrated visibility data from both bands were combined, averaged to 32\,MHz channels, and imported into DIFMAP \citep{1997ASPC..125...77S}. 
Bright field sources were modeled separately for each band using the visibility data and a combination of point-source and Gaussian components with power-law spectra. After subtracting the modeled field sources from the visibility data, \object{GW170817} dominates the residual image. Restored naturally-weighted images for each band were generated by convolving the restoring beam and modeled components, adding the residual map and averaging to form a wide-band image. Image-based Gaussian fitting with unconstrained flux density and source position was performed in the region near \object{GW170817}.
% The ATCA flux densities of GW170817, after applying a correction factor of 1.25 (see \S\ref{subsec:spectral} for details), are reported in Table~\ref{tab:radiodata}.

\subsection{uGMRT} \label{subsec:gmrt}

We observed GW170817 with the uGMRT in Band 3 (effective bandwidth 550--750 MHz) (PI: Mooley). The observations were carried out with 400 MHz correlator bandwidth centered at 750 MHz using the non-polar continuum interferometric mode of the GMRT Wideband Backend \citep[GWB;][]{reddy2017}. 
The epochs from 2018 May and 2018 Jun were divided into several short ($\sim$1--3 hr) observations carried out over several days. 
3C286 was used as the absolute flux scale and bandpass calibrators, while phase calibration was done with 3C283. 
These data were calibrated and RFI flagged using a custom-developed pipeline in CASA. 
The data were then imaged interactively with the CASA task CLEAN, while incorporating a few iterations of phase-only self-calibration and one amplitude+phase self-calibration step.
% The uGMRT flux densities of GW170817 are reported in Table~\ref{tab:radiodata}.

\subsection{MeerKAT} \label{subsec:meerkat}

We observed GW170817 with the new MeerKAT telescope \citep{camilo2018,2016mks..confE...1J} on 2018 Mar 02, Apr 26 and May 05. The first of these observations used 16 antennas and the ROACH2 correlator; the latter used 61 antennas and the SKARAB correlator. All observations were made at L-band, covering 900--1670 MHz. After flagging for RFI, the effective bandwidth used was 486 MHz.
PKS\,1934$-$638 was used as the flux and bandpass calibrator and for initial phase reference.
Data processing was done with the MeerKATHI pipeline (Makhatini et al. in prep).
To correct for the uncertainties in the frequency-dependent primary beam correction and pointing errors, we used direction-dependent gain calibration for bright sources spread across the $\sim1$ deg field of view.

\section{Results} \label{sec:results}
\subsection{Spectral analysis}\label{subsec:spectral}

In \cite{2018ApJ...858L..15D} we studied the radio spectral evolution of GW170817, finding that over the first 120 days the radio spectral index was constant with a value of $\beta=-0.57\pm 0.04$. This value of $\beta$ is fully consistent with the value derived by \citet{2018arXiv180502870A}, $\beta=-0.74\pm 0.2$, at 217 days. Here and elsewhere in the paper we characterize the flux density evolution of the light curve using $F_\nu(t,\nu)\propto{t^{\alpha} \nu^{\beta}}$ where $\alpha$ and $\beta$ are the temporal and spectral power-law indices, respectively. 

We have used the new data to look for changes in the radio spectral index. Specifically we have searched for a steepening of the spectral index of order $\Delta\beta=0.5$ as expected if the synchrotron cooling break had moved through the radio band on timescales of 200--300 days \citep{1998ApJ...497L..17S}. We compare our radio measurements from May 2018 reported in Table~\ref{tab:radiodata} with the X-ray measurement from 2018 May 3--5 \citep{2018arXiv180502870A,nynka2018,2018arXiv180806617V} to derive a spectral index of $\beta_{\rm XR}=-0.56\pm0.01$. Our most precise (simultaneous) two-point measurement of the radio-only spectral index is with the VLA data between 3--10 GHz in Mar 2018 (day$\sim$220), $\beta=-0.52\pm0.09$. 
Both these measurements are consistent with the precise radio-to-X-ray spectral index $\beta=-0.584\pm0.006$ at 160 days post-merger derived by \cite{2018ApJ...856L..18M}.
We see no evidence for a spectral steepening, and can rule out the presence of a cooling break for which we expect $\beta\sim-1.1$ \citep{2018arXiv180502870A}.

\begin{figure}
\centering
\includegraphics[width=3.5in]{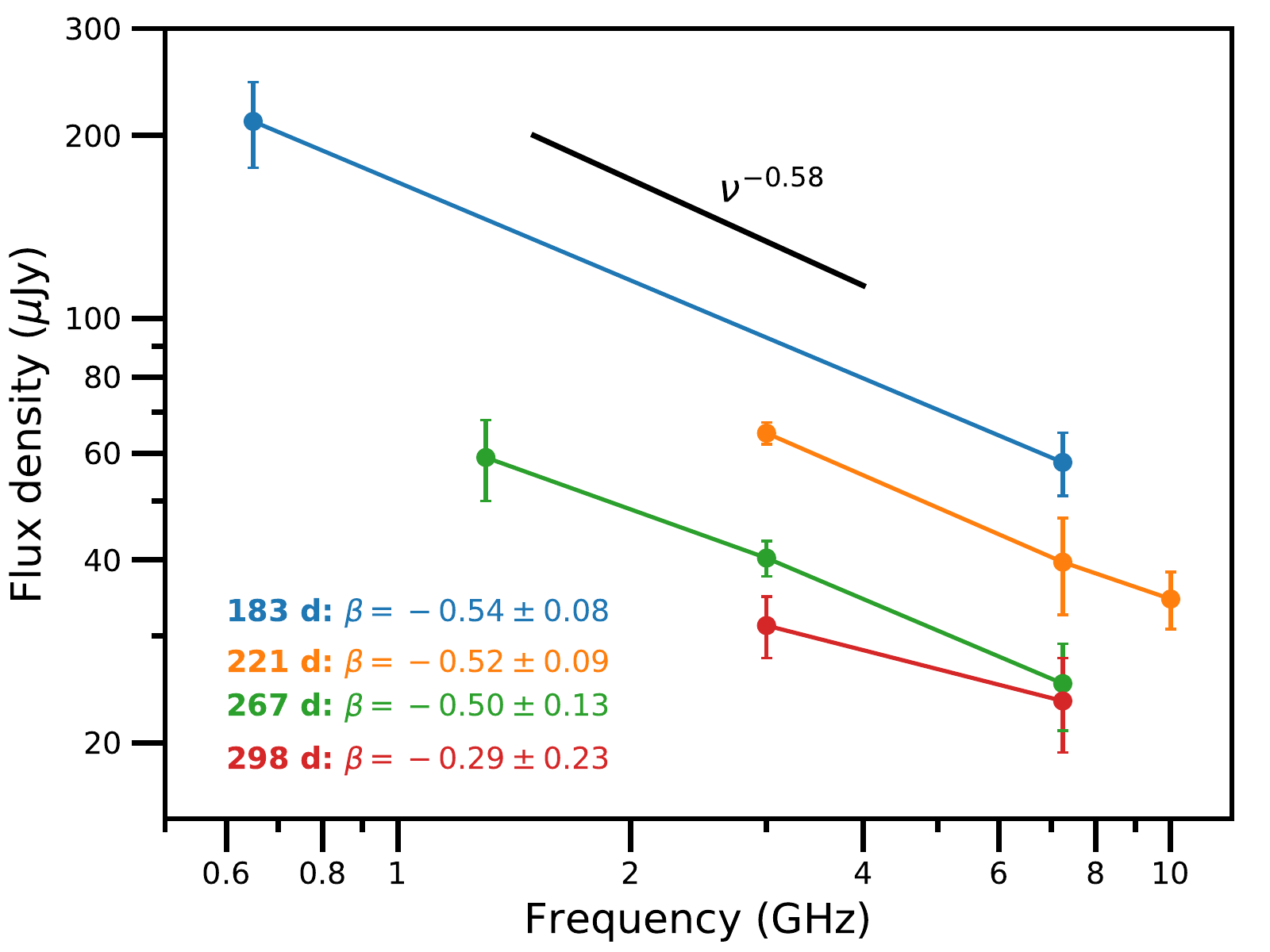}\\
\caption{Radio spectral indices between 0.6--10 GHz spanning four epochs observed during the decline part of the light curve. The different epochs are color coded. The approximate number of days post-merger and the corresponding spectral indices are given in the legend. A joint fit to all the radio data implies $\beta=-0.53\pm0.04$ (see \S\ref{subsec:light curve}), in good agreement with the radio to X-ray spectral index measurements (shown by the black $\nu^{-0.58}$ line). With this measurement, we can rule out any spectral steepening expected, for example, due to the  presence of a cooling break in or between the radio/X-ray bands.}
\label{fig:spec}
\end{figure}

In these spectral comparisons between telescopes, we see some evidence for a scaling offset in the late-time ATCA data. While the in-band 5.5--9 GHz spectral indices derived for the ATCA data are consistent with $-0.584$ broadband value above, the flux densities are higher by about 25\% when the data are compared with contemporaneous measurements made with the VLA, MeerKAT and uGMRT.
% but there is an excess of about 25\% when the data (we consider only 7.25 GHz flux density values here) are compared with contemporaneous measurements made with the VLA, MeerKAT and uGMRT. 
The origin of excess in the ATCA data is currently being investigated\footnote{This appears to be a systematic offset. The origin of the offset is not in the flux scale, since the radio spectrum of the phase reference source is consistent between all the epochs from the different telescopes. 
% As the ATCA array was in its more compact configurations when these observations were taken, it may be due to host galaxy contamination.
The ATCA array has a more compact configuration, so it may be due to host galaxy contamination.}.
In Table~\ref{tab:radiodata} we have reported the corrected ATCA flux densities. 
The radio-only spectral index measurements from four epochs observed during the decline of the light curve are shown in Figure~\ref{fig:spec}.
The spectral index between the 0.65 GHz uGMRT and 7.25 GHz ATCA data \citep[corrected from the value reported in ][]{2018ApJ...858L..15D} obtained around 2018 Feb 16 (183 d post-merger) is $-0.54\pm0.08$.
Below, we describe a joint fit to multi-frequency light curve data, including the spectral and temporal indices. 

% We derive the spectral index between 3 GHz and 7.25 GHz of $\beta=-0.3\pm 0.2$. This value of $\beta$ is less well-determined than the $\beta$ values above which were measured when \object{GW170817} was brighter. However, it is sufficiently accurate to 
% likewise  
% From here on we will use the more accurate radio-to-X-ray spectral index $\beta=-0.584\pm0.006$ at 160 days post-merger derived by \citet{2018ApJ...856L..18M}.

%We first re-visit the spectral behavior of the radio emission.As in \citet{2018Natur.554..207M} we fit a power-law of the form $S_\nu \propto \nu^{\alpha}t^{\delta}$ to the first 120 days of the radio light curve (before any sign of a turnover) and find a spectral index $\alpha=-0.57\pm 0.04$ and temporal index $\delta=0.84\pm0.05$. This is consistent with \citet{2018Natur.554..207M} and with \citet{2018arXiv180103531M}, who find a joint radio-to-X-ray spectral index $\alpha=-0.585\pm0.005$ at 110 days and $\alpha=-0.584\pm0.006$ at 160 days post-merger.

%We examined the variability of the spectral behavior using all quasi-simultaneous radio observations.  We identified data-sets with more than one observation within $\pm1\,$day and fit for a spectral index. These values are shown in Figure~\ref{fig:alpha_timeseries}. We find the data largely consistent with a constant spectral index, with $\chi^2=15.9$ for 12 degrees-of-freedom.  There appears to be no evidence for significant change in the spectrum of the source, consistent with previous radio, X-ray and \textit{HST} observations \citep{2018arXiv180106164D,2018arXiv180102669L,2018Natur.554..207M,2018arXiv180103531M,2018arXiv180302768R}. 
%SO FAR WE HAVE NOT DONE A JOINT FIT OF THE VLA AND ATCA DATA TO FIND THE SPECTRAL INDEX. CURRENTLY, WE FIND THAT BETWEEN 3 GHz and 7.25 GHz, THE SPECTRAL INDEX IS -0.3+/-0.2 (compare with -0.585+/-0.003 from Marutti et al. 2018).

\subsection{Light curve analysis} \label{subsec:light curve}

We began by performing a joint fit to 
% {\it all} the 
radio data published 
% to date
till day 300 post-merger. This includes the data in Table~\ref{tab:radiodata} together with \cite{2017Sci...358.1579H,2017ApJ...848L..21A,2018Natur.554..207M,2018ApJ...856L..18M,2018arXiv180302768R,2018ApJ...858L..15D}. We used a smoothly-broken power law model, incorporating the frequency dependence, of the form $2^{1/s}\nu^{\beta}F_{\nu,p}~(t^{-s\alpha_1} + t^{-s\alpha_2})^{-1/s}$ \citep{beuermann1999,2018arXiv180502870A}. Here, $\nu$ is the observing frequency normalized to 3 GHz, $F_{\nu,p}$ is the flux density at 3 GHz at the time of light curve peak, $t$ is the time in units of the time to light curve peak ($t_p$), $s$ is the smoothness parameter, and $\alpha_1$,$\alpha_2$ are the power-law rise and decay slopes. This Markov chain Monte Carlo (MCMC) fitting was done\footnote{We chose 100 walkers, 1000 steps and flat priors on all of the parameters.} using the Python package {\tt emcee} \citep{foreman-mackey2013}.  We obtain best-fit values of $F_{\nu,p}=118^{+14}_{-7}$ $\mu$Jy, $t_p=167^{+14}_{-7}$ days, $\alpha_1=0.80^{+0.06}_{-0.05}$, $\alpha_2=-2.16^{+0.23}_{-0.61}$, log$_{10}(s)=0.59^{+0.77}_{-0.37}$ and $\beta=-0.61^{+0.03}_{-0.07}$ (68\% confidence interval, i.e. 1$\sigma$; see Table~\ref{tab:params}). We also introduced a scale factor into the MCMC fit to explore a possible 25\% offset in the ATCA flux densities suggested by the spectral fits in \S\ref{subsec:spectral}. We find that a scaling factor of $\sim$20\% is slightly preferred over unity\footnote{Median flux multiplication factor is 0.83 and the 68\% confidence interval is 0.75--1.07. Note that the scaling factor is required for all ATCA data (reported here and previously). As an experiment, we have also performed a fit without including an ATCA flux scaling factor in the MCMC analysis, and the $\chi^2$ is significantly worse in this case as expected (87.4 versus 67.4). Nevertheless, we get $\alpha_2=1.86^{+0.17}_{-0.23}$ without the scaling factor.}.

\setcounter{table}{1}
\begin{table*}
\centering
\caption{Parameters obtained from fitting a smoothly broken power law model to the radio light curve}
\label{tab:params}
\begin{tabular}{lllllll}
\hline\hline
No. & F$_{\nu,p}$ & $t_p$ & $\alpha_1$ & $\alpha_2$ & log$_{10}$(s) & $\beta$\\
     & ($\mu$Jy) & (d) \\
\hline
1   & $118^{+14}_{-7}$ & $167^{+14}_{-7}$ & $0.80^{+0.06}_{-0.05}$ & $-2.16^{+0.23}_{-0.61}$ & $0.59^{+0.77}_{-0.37}$ & $-0.61^{+0.03}_{-0.07}$ \\
% 2   & $99\pm9$  & $172^{+9}_{-6}$ & $0.79^{+0.05}_{-0.04}$ & $-2.38^{+0.26}_{-0.44}$ &   $0.67^{+0.52}_{-0.32}$ & $-0.54\pm0.04$ \\
2   & $98^{+8}_{-9}$  & $174^{+9}_{-6}$ & $0.78\pm0.05$ & $-2.41^{+0.26}_{-0.42}$ &   $0.70^{+0.49}_{-0.34}$ & $-0.53\pm0.04$ \\
% 3*   & $117\pm7$  & $164^{+8}_{-9}$ & $0.81^{+0.07}_{-0.06}$ & $-2.14^{+0.24}_{-0.28}$ & \ldots & \ldots\\
3*   & $120\pm9$  & $164\pm7$ & $0.83\pm0.07$ & $-2.23\pm0.24$ & \ldots & \ldots\\
\hline
\multicolumn{7}{p{4.5in}}{Row 1 gives the joint fit for the radio data reported here together with that reported previously, row 2 gives the fit for the radio data reported here and together with previous data referenced robustly with our method of flux determination, and *row 3 gives the fit (broken power law) for the 3 GHz VLA-only data. Column descriptions: $F_{\nu,p}$ ($\mu$Jy) is the flux density at 3 GHz at the time of light curve peak, $t_p$ is the time of peak (days post-merger), $\alpha_1$,$\alpha_2$ are the power-law rise and decay slopes, $s$ is the smoothness parameter, and $\beta$ is the spectral index. See \S\ref{subsec:light curve} for details.}
\end{tabular}
\end{table*}

\begin{figure}
\centering
\includegraphics[width=3.5in]{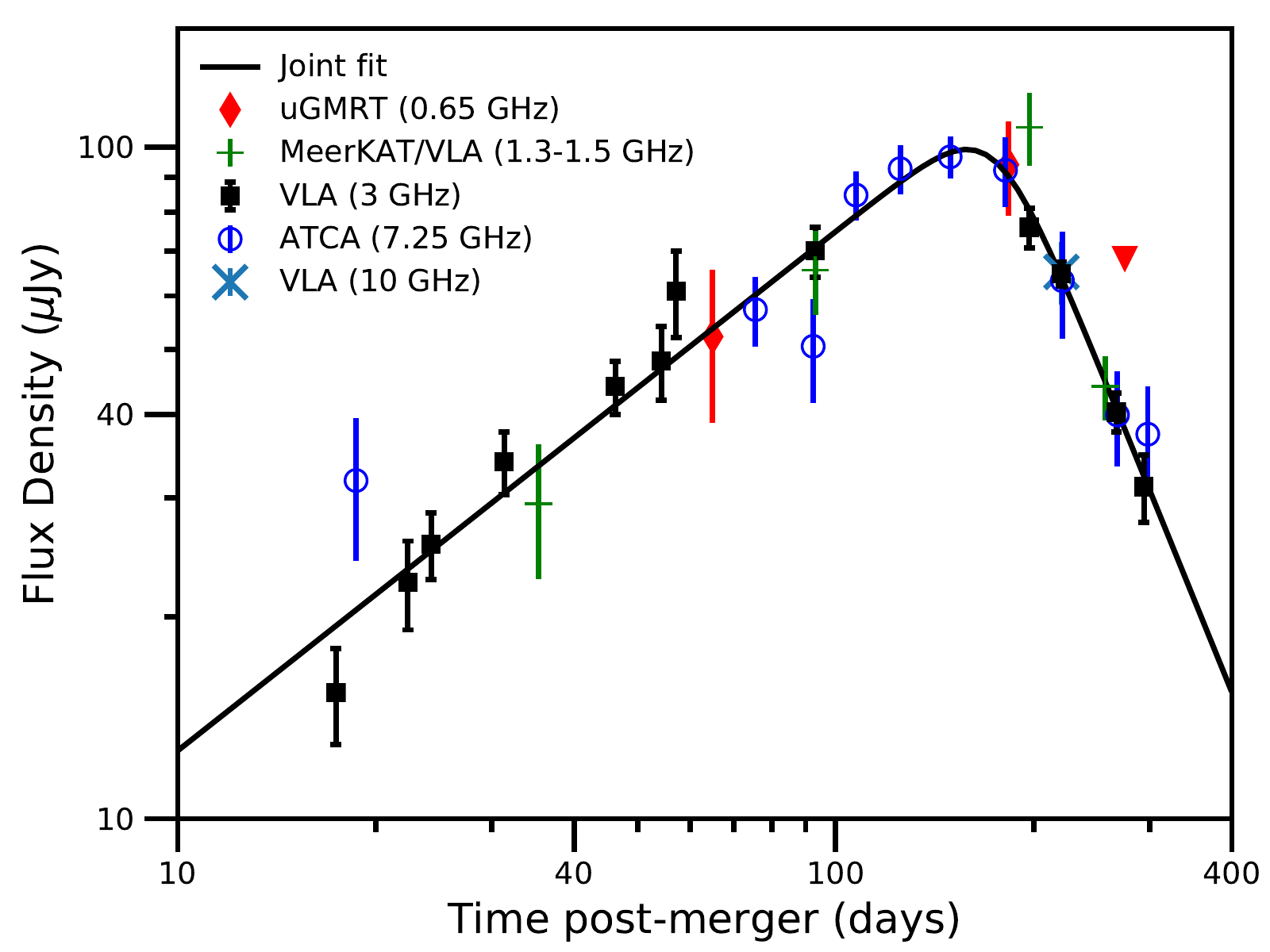}\\
\caption{The radio light curve of \object{GW170817} spanning multiple frequencies, and scaled to 3 GHz using the spectral index ($\nu^{-0.53}$) derived from our MCMC analysis. The data from the VLA (filled black squares for 3 GHz and green crosses for 1.5 GHz),  the ATCA (blue circles), the MeerKAT (green crosses) and the uGMRT (red diamonds for detections and triangle for upper limit) are as reported in Table 1. We also include the data at 0.65 GHz, 1.5 GHz, 3 GHz, and 7.25 GHz reported previously \citep{2017Sci...358.1579H,2018Natur.554..207M,2018ApJ...858L..15D}. Our best-fit smoothed broken power-law model to all these data (see \S\ref{subsec:light curve}) is shown as a solid curve. The power-law decline index obtained is $-2.4^{+0.3}_{-0.4}$. For comparison, a broken power-law fit to the 3 GHz VLA-only data gives $-2.2\pm0.2$. Both fits are thus consistent with $t^{-p}$ decline in the light curve, where $p$ is the electron power-law distribution index. 
% We see that the transition from the $t^{0.8}$ power law rise to $t^{-p}$ decline in the light curve takes place rapidly, over a period of 25--105 days (see \S\ref{subsec:light curve}).
}
\label{fig:lc}
\end{figure}

\begin{figure*}
\centering
\includegraphics[width=7in]{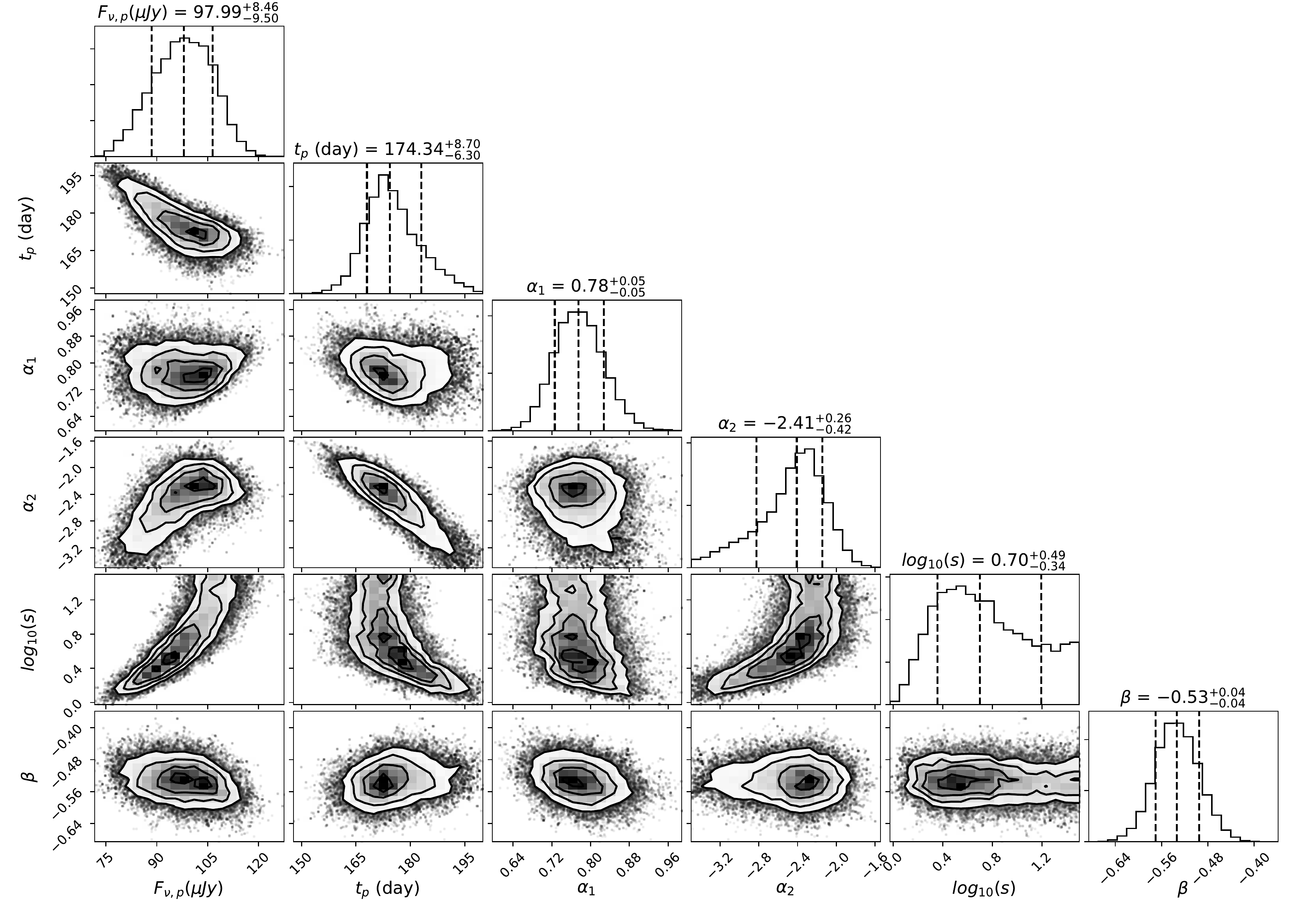}
\caption{A corner plot showing the results of our MCMC fitting of the radio light curve (Figure~\ref{fig:lc}) using the VLA, ATCA, MeerKAT and uGMRT data. $F_{\nu,p}$ is the flux density at 3 GHz at the time of light curve peak, $t_p$ is the time to light curve peak, $\alpha_1$,$\alpha_2$ are the power-law rise and decay slopes, $s$ is the smoothness parameter, and $\beta$ is the spectral index. See \S\ref{subsec:light curve} for details. In each contour plot and histogram, the 16, 50, 84 percentiles are marked.}
\label{fig:corner}
\end{figure*}

Next we fit only the data in Table~\ref{tab:radiodata} together with previous data at 0.65 GHz, 1.5 GHz, 3 GHz and 7.25 GHz referenced robustly with our method of flux determination \citep[][flux density values given in Table~\ref{tab:oldradiodata}]{2017Sci...358.1579H,2018Natur.554..207M,2018ApJ...858L..15D}.
% and two measurements at 3 GHz close to the peak of the light curve. 
% that we have reported previously. We exclude the 3 GHz data from \citet{2018arXiv180502870A} 
% We also included two 3 GHz data points from \cite{2018ApJ...856L..18M}, obtained close to the light curve peak, to get better sampling of the transition between the power-law rise and power-law decline.
Our best-fit values are given in Table~\ref{tab:params}, and are consistent with the fit using all of the data above.  In particular we find $\alpha_2=-2.4^{+0.3}_{-0.4}$.
% We obtain best-fit values of $F_{\nu,p}=99\pm9$ $\mu$Jy, $t_p=172^{+9}_{-6}$ days, $\alpha_1=0.79^{+0.05}_{-0.04}$, $\alpha_2=-2.4^{+0.3}_{-0.4}$, log$_{10}(s)=0.7^{+0.5}_{-0.3}$ and $\beta=-0.54\pm0.04$.
% (68\% confidence interval, i.e. 1$\sigma$). % The smoothness parameter therefore lies between 1--16 (1$\sigma$ range), implying that the transition from $\alpha_1$ to $\alpha_2$ is fairly sharp (see below). 
Figure~\ref{fig:lc} shows the multi-frequency radio data scaled to 3 GHz, and the joint fit to these data (solid line).
% Inspecting the light curve by-eye, we find that the transition from the power-law rise to the power-law decline takes place over a period of $<$80 days.
Figure~\ref{fig:corner} shows the corner plot with the results of the MCMC fit. 

By taking the limit in which the $t^{-s\alpha_1}$ term dominates\footnote{We derive the time at which one term dominates over the other by a factor of $\sim$20. The quoted time values are the median of the distributions and their 16 and 84 percentiles are quoted as the uncertainties.} over the $t^{-s\alpha_2}$ term in the smoothly-broken power law expression given above, we derive that the transition from the power law rise to the power law decay takes place between 
% $156^{+13}_{-17}$ and $183^{+36}_{-16}$ days
$158^{+13}_{-18}$ and $183^{+42}_{-15}$ days post-merger, i.e. over a timescale of $24^{+58}_{-24}$ days.
This implies that the transition from $\alpha_1$ to $\alpha_2$ is fairly sharp, possibly taking place over a small fraction of the time taken to reach the light curve peak. We return to this point in \S\ref{sec:discussion}.

%By performing a joint fit to all the radio data published to date, we get fit parameters that are consistent with (but somewhat less well-constrained than) the parameters derived above, e.g. $\alpha_2=-2.16^{+0.27}_{-0.61}$. However, since there may be systematic uncertainties involved in the calibration across data taken from different telescopes and obtained at different frequencies, we independently fit our 3 GHz VLA-only data. In this case, the light curve is too sparsely sampled to be able to fit for the smoothness parameter, and hence we use a simple broken power law model (this corresponds to $s \rightarrow \infty$) instead. We obtain best fit parameters (at 3 GHz) of $F_{\nu,p}=117\pm7$ $\mu$Jy, $t_p=164^{+8}_{-9}$ days, $\alpha_1=0.81^{+0.07}_{-0.06}$ and $\alpha_2=-2.14^{+0.24}_{-0.28}$. The decline is somewhat shallower than, but in good agreement with, the smoothly-broken power law model parameters. With less data and a shorter time-baseline \cite{2018ApJ...858L..15D} derived $\alpha_2=-1.6\pm0.2$, which is similar to the value that \citet{2018arXiv180502870A} find, $\alpha_2=-1.6^{+0.2}_{-0.3}$. 

The reduction in the uncertainties for $\alpha_2$ in the second fit hints that there may still be systematic uncertainties involved in the calibration across data taken from different telescopes and obtained at different frequencies. Thus we chose to independently fit the 3 GHz VLA-only data as was first done in \citep{2018arXiv180609693M}. In this case, the light curve is too sparsely sampled to be able to fit for the smoothness parameter, and hence we use a simple broken power law model (this corresponds to $s \rightarrow \infty$) instead. 
% We obtain best fit parameters (at 3 GHz) of $F_{\nu,p}=117\pm7$ $\mu$Jy, $t_p=164^{+8}_{-9}$ days, $\alpha_1=0.81^{+0.07}_{-0.06}$ and $\alpha_2=-2.14^{+0.24}_{-0.28}$. 
Table~\ref{tab:params} gives the parameter values from the fitting, and we find that $\alpha_2=-2.2\pm0.2$.
The decline is somewhat shallower than, but in good agreement with, the smoothly-broken power law model parameters. The remaining parameters such as the slope of the rise, the peak flux density and the time of peak all agree well with each other and with previous fits in the literature. The main point here is that our key results are robust to different choices of the data that we used in the fit. 
% The fit parameters derived for the different data choices are given in Table~\ref{tab:params}.

Summarizing, we measure a sharp transition of the afterglow light curve of GW170817 about 170 days post-merger with a steep power-law slope of $\alpha_2=-2.2$. The result confirms our earlier determination of $\alpha_2$ first reported in \citet{2018arXiv180609693M}. With less data and a shorter 
time-baseline \cite{2018ApJ...858L..15D} derive a more shallow decay index $\alpha_2=-1.6\pm0.2$, which is similar to the value that \citet{2018arXiv180502870A} find, $\alpha_2=-1.6^{+0.2}_{-0.3}$. Our more precise values of $\alpha_2$ lie within the 68\% confidence interval of \citet{2018arXiv180806617V} but we measure a larger value for the smoothness parameter.

\section{Discussion}\label{sec:discussion}
%We have presented new ATCA and VLA observations of \object{GW170817} covering the period 125--200\,days post-merger. Combined with previous radio observations these data show no evidence for spectral evolution, but they conclusively show that the radio counterpart has peaked in brightness at $149\pm2$\,days post-merger and is currently declining.  We use this to rule out emission being caused by highly energetic, quasi-isotropic outflow or highly energetic, highly-relativistic outflow  but are not able to uniquely determine the geometry and structure of the actual outflow material. Continued radio monitoring will allow the temporal decay index to be accurately determined, although this may not be sufficient to establish the presence of a successful jet \citep{2018arXiv180109712N} and degeneracies in the ejecta total energy and the density of the circum-merger environment may preclude confirmation of any particular model. Polarization measurements and VLBI observations should be able to break this degeneracy and thus distinguish between the models \citep[also see][]{2018arXiv180305892G}.

% Say why is this important to distinuish between models. What are the implications.

Before interpreting the light curve of \object{GW170817} directly, it is illustrative to review the two asymptotes of late-time light curve behavior from afterglow models. Afterglow spectra and the light curves of GRBs have long been used to infer the geometry and dynamical state of the ejecta \citep[e.g.][]{galama1998,1999ApJ...523L.121H}. For a spherical relativistic fireball the flux density will decline as $F_\nu\propto t^{\alpha} \nu^{\beta}$, in which $\alpha=-3(p-1)/4$ and $\beta=-(p-1)/2$ when the observing frequency $\nu_\circ$ is below the synchrotron cooling break $\nu_c$, and $\alpha=-(3p-2)/4$ and $\beta=-p/2$ when the cooling break lies below the observing frequency \citep{1998ApJ...497L..17S}. Here $p$ is the usual power-law index for the energy of the accelerated electrons ($p>2$). For a jet viewed on-axis at late times the power-law decay index is $\alpha=-p$, independent of whether the cooling break is above (i.e. $\beta=-(p-1)/2$) or below (i.e. $\beta=-p/2$) the observing frequency \citep{1999ApJ...519L..17S}.

For \object{GW170817} we show here (\S{\ref{subsec:spectral}}), as also demonstrated elsewhere 
% and elsewhere 
\citep{2018ApJ...858L..15D, 2018arXiv180502870A}, that in the radio regime $\beta$ is consistent with the value found from fitting across the radio and X-ray regimes, $\beta=-0.584$, and that $\nu_c$ lies well above the radio (and likely also the X-ray) band. Thus $\beta=-(p-1)/2$ and $p=2.17$, for which we expect the late-time power-law decay index $\alpha$ to lie between $-0.88$ (i.e. quasi-spherical, cocoon-dominated) and $-2.17$ (jet-dominated).  Eventually we also expect the outflow to become non-relativistic and this can give rise to an achromatic change in the GRB afterglow light curves. Dynamical transitions to the non-relativistic phase have been claimed for both spherical and jet-like outflows \citep{2000ApJ...537..191F,2005ApJ...619..994F,2008A&A...480...35V}. The timescale on which this occurs is approximately when the rest mass energy of the material swept up by the shock is comparable to the kinetic energy of the outflow \citep{2000ApJ...537..191F}. The side-ways expansion of the jet becomes important and eventually the outflow becomes quasi-spherical \citep{2000ApJ...537..191F}. At this time $\alpha_{\rm nr}=-(15p-21)/10$ for $\nu_\circ<\nu_c$ and $\alpha_{\rm nr}=-(3p-4)/2$  for $\nu_\circ>\nu_c$ \citep{2000ApJ...538..187L}. Thus for the nominal parameters of \object{GW170817}, a spherical outflow undergoing a non-relativistic transition would be expected to show an achromatic {\it steepening} while for a jet the light curve would {\it flatten}, both with a value of $\alpha_{\rm nr}=-1.15$. 
%(estimate the approximate NR timescale from Frail et al 2000 paper and the energy+density from GW 170817 models).

The afterglow light curves of jet-like outflows are altered by observing them at different viewing angles away from their symmetry axis. In this case the structure of the ejecta becomes important. Early work investigated the role of viewing angle for simple uniform or top-hat jets \citep{2000ApJ...538..187L} and jets with azimuthal structure \citep[e.g.][]{2002MNRAS.332..945R,2003ApJ...591.1075K}. More recent modeling has considered structure jets whose ejecta have both azimuthal and radial structure \citep{2018MNRAS.473L.121K,2018MNRAS.473..576G,2018arXiv180409345X,2018PhRvL.120x1103L}. The effects of jet structure and viewing angle are most pronounced at or near the peak of the light curve. These models generally predict a slow temporal evolution of $\alpha$, with the break between the rise and the decay taking place over a significant fraction of the peak time \citep{2003ApJ...591.1086G,2003ApJ...592..390P,2018MNRAS.tmp.1159G,2018arXiv180603843L}. However, at late times all of these off-axis light curves models approach the behavior of an on-axis jet where the slope of the temporal index, as noted earlier, is $\alpha=-p$.

%Preliminary, stream of conciousness
While the predicted late-time light curves of afterglows exhibit a diverse range of behaviors, the observed decay of \object{GW170817} is remarkably simple. A single power-law with $\alpha_2=-2.4^{+0.3}_{-0.4}$ ($2.2\pm0.2$ for the 3 GHz VLA-only data) fits all the data post-peak. This power-law index is a clear signature of a relativistic jet. This is a strong jet-dominated outflow, i.e. there is no support for intermediate slopes as might be expected if a quasi-spherical cocoon was contributing to the emission. Likewise, we see no evidence for a spectral change due to synchrotron cooling (\S\ref{subsec:spectral}) nor do we see a dynamical transition to non-relativistic motion that would manifest itself by an achromatic break in the light curve. Another important feature of the light curve in Figure \ref{fig:lc} is the sharpness of the transition from a power-law rise to decay. The change from $t^{0.8}$ to $t^{-p}$ takes place over $24^{+58}_{-24}$ days, a result that appears to be at odds with the predicted temporal evolution of $\alpha$ for current off-axis, structured jet models \citep{2003ApJ...591.1086G,2003ApJ...592..390P,2018MNRAS.tmp.1159G}. More detailed modeling of GW170817 is needed to see whether structured jet-like outflows can reproduce this sharp transition.

The sharpness of the peak and the slope of the power-law decline depend on the viewing/observing angle, $\theta_v$, and jet core half-opening angle\footnote{Since the energy distribution at the core is expected to be roughly uniform we approximate its contribution as being generated by a top-hat jet with half-opening angle, $\theta_j$.}, $\theta_j$, and more specifically on the ratio between them. In order to constrain this ratio we consider only the contribution from the core of the jet, which dominates the emission near the peak and during the decay. 
Thus, while the rising part of the light curves that we calculate does not fit the observations, the peak and the decay should. Using this approximation, we can now derive constraints on $\theta_j/\theta_v$ that provides the observed  transition from the peak of the light curve and the steep decline.
% within a factor of ?? in time.  

We make a rough analytic approximation. The peak occurs approximately when we start seeing the near edge of the jet core, and the $t^{-p}$ power-law decline begins roughly when the jet centroid comes into view. The sharpness of the light curve peak and the immediate transition to $t^{-p}$ decline implies that we are in the regime\footnote{If $(\theta_v-\theta_j)\lesssim\theta_j$ there will be a long-lived phase during which the light curve decays as $t^{-1}$. } $\theta_v-\theta_j \gg \theta_j$. We denote by $t_1$ as the time that we see the edge of the jet, namely $\Gamma(t_1) \simeq 1/(\theta_v-\theta_j)$,  and $t_2$ as the time that we see the jet axis, $\Gamma(t_2) \simeq 1/\theta_v$. Now, ignoring sideways spreading of the jet we can approximate $\Gamma \propto t^{-3/8}$ \citep[e.g.][]{1998ApJ...497L..17S} to obtain      
$\Delta t / t = (t_2-t_1)/t_2 \simeq [\theta_v^{8/3} - (\theta_v-\theta_j)^{8/3}] / \theta_v^{8/3} \simeq (8/3) \theta_j/\theta_v$. Here we use the approximation that $\theta_j/\theta_v$ is much smaller than unity. Observationally, $t_1$ occurs sometime during the transition from the $t^{0.8}$ rise to the peak of the light curve and $t_2$ occurs when the decay phase approaches $t^{-p}$. Using the results from our MCMC analysis, we find that $\Delta t / t \lesssim 1/3$ (68\% confidence or better, depending on the where $t_1$ and $t_2$ lie), indicating that $\theta_v \gtrsim 8\theta_j$. 
% lies between $\sim$1--0.25. Solving for $\theta_j/\theta_v$ we get permissible values between 5--20.
To verify this simple estimate we produced light curves from a top-hat jet using semi-analytic code\footnote{The semi-analytic code \citep{soderberg2006} takes proper account of all the relativistic effects such as Lorentz boost and light arrival time from each point in the jet and it includes lateral spreading of the jet.  Following corollary comparison of the code results to BOXFit \citep{vanEerten2012} we use lateral spreading of 30\% of the local thermal speed. We verify that varying the spread velocity, including taking no spreading at all does not change the result significantly.}. We find that light curves that fit the transition seen from the peak to the decline of the light curve have  $\theta_v \simeq 6\theta_j$. 
Together with the upper limit on the observing angle from the LIGO-Virgo Collaboration \citep[$\theta_v\lesssim 28^o$;][]{2017PhRvL.119p1101A}, our constraint implies $\theta_j\lesssim5^o$.
If instead we use the estimate of $\Gamma\simeq4.1\pm0.5$ close to the peak of the light curve from the VLBI measurement \citep{2018arXiv180609693M}, then we get $\theta_j\lesssim3^o$ and $\theta_v\simeq15^o$.

\section{Summary \& Conclusions}\label{sec:conclusion}

In this companion paper to our polarization and high angular resolution imaging studies, we have presented deep VLA, ATCA, MeerKAT and uGMRT observations (\S\ref{sec:data}) of GW170817 post-peak of the light curve, i.e. between 180--300 days post-merger. Our spectral analysis does not yield any evidence for the cooling break to have entered the radio band, or of any achromatic steepening in the light curve indicating the transition of the outflow into the non-relativistic regime (\S\ref{subsec:spectral}). We find that the light curve decay is consistent with $t^{-2.2}$ (\S\ref{subsec:light curve}), which is a classic signature of a jet where the slope of the decay is equal to the power-law energy index $p$ of the synchrotron-emitting electrons. We also find that the transition from the power-law rise to decay ($t^{0.8}$ to $t^{-2.2}$) is fairly sharp. 
% , and happens within a span of $27^{+79}_{-1}$ days.

The data on the light curve decline reported previously \citep{2018ApJ...858L..15D, 2018arXiv180502870A} has not been sufficiently constraining to unambiguously distinguish between cocoon-dominated and jet-dominated emission.
The new data that we have reported here securely implies a decline in the light curve consistent with $t^{-p}$ (where $p$ is the electron energy distribution power law index), which is a strong evidence for the late-time afterglow being jet-dominated. Our observations support recent claims from hydrodynamic modeling that \object{GW170817} produced a successful jet \citep{2018arXiv180610616D}. Using these new data we can also derive robust constraints on the smoothness parameter ($s$) and therefore the sharpness of the light curve peak, something which has not been possible with previously-reported data. 
Together with the sharpness of the peak, the steep decline indicates that the jet is extremely narrow and that most of the outflow energy of GW170817 resides in the jet. 
Through simple analytical arguments we are able to place a constraint on the geometry, $\theta_v \gtrsim 8\theta_j$ ($\theta_v \gtrsim 6\theta_j$ with semi-analytical modeling; $\theta_v$ is the viewing angle and $\theta_j$ is the jet half-opening angle), and implies $\theta_j\lesssim5^o$ if we further use the viewing angle constraint provided by the LIGO-Virgo Collaboration.
Using $\Gamma\simeq4$ close to the peak of the light curve (estimated from the observed superluminal motion in GW170817) gives $\theta_j\lesssim3^o$ and $\theta_v\simeq15^o$.

These conclusions are consistent with results from VLBI and hydrodynamical simulations \citep{2018arXiv180609693M}, from parametric modeling of the jet \citep{hotokezaka2018,2018arXiv180800469G,2018arXiv180806617V}, and from polarization \citep{2018ApJ...861L..10C}. The jet opening angle for GW170817 appears to be somewhat smaller than the median found for short GRBs, $<\theta_j>\simeq16^o\pm10^o$ \citep[$\theta_j$ estimates for bursts displaying jet breaks lie between $\sim2^o-8^o$; e.g.][]{fong2015}, but is consistent with the estimates for short GRBs like 090510 and 150101B \citep[][see also \cite{2018arXiv180806617V}]{kumar2010,nicuesa2012,troja2018c} and the tail end of the long GRB distribution \citep[e.g.][]{goldstein2016}. With the confirmation of the successful jet in GW170817, our polarization upper limit from \cite{2018ApJ...861L..10C} implies high isotropy of the magnetic field.
% is highly random within the plane of the shock occurring between the jet and the interstellar medium. 

% Most important paragrapH: put it all together from early to late times and what does it all mean for the AG of GW 170817. Early times. Energy injection, coocoon dominated. late times jet expanding laterally into the los. There was a jet, it did survive. geomety of not as simple as initially suggested. SHB and GW are linked but more complex if GW 170817 is the prototype. 
At early-times ($\simeq$100 days), the steady rise in the light curve indicated that there was continuous energy injection within the outflow emitting in our line of sight \cite[e.g.][]{nakar-piran2018}, and implied the presence of a mildly relativistic wide-angle outflow with significant angular and/or radial structure (consistent with a cocoon-dominated outflow). 
At later times, the successful narrow jet came into view and dominated the late-time afterglow. The confirmation of a jet strengthens the link between neutron star mergers and regular SHBs.

Like some previous long-duration GRBs \citep[e.g.][]{2003Natur.426..154B}, \object{GW170817} has given us insights into the 
structure of the wide-angle outflows surrounding the jet core, and the simple top-hat jet for SHBs will likely have to be revised \citep{2017ApJ...834...28N, 2017MNRAS.471.1652L,2017ApJ...848L...6L}. The confirmation of the wide angle outflow, that dominated the outflow at early times, bodes well for the future detection of the EM counterparts of GW sources observed at larger viewing angles.

\begin{table}
\caption{Previous radio observations of \object{GW170817}}
\label{tab:oldradiodata}
\begin{tabular}{llllll}
\hline
\hline
UT date & $\Delta$T$^{\dag}$ & Telescope & $\nu$ & $F_\nu$ & $\sigma_\nu$\\
        & (d)                &           & (GHz) & ($\mu$Jy) & ($\mu$Jy)\\
\hline
2017 Sep 02--04   & 17.4     & VLA       & 3     &  15.4    & 2.5 \\
% 2017 Dec 10       & 115.1              & VLA       & 3     &  92.8    & 9.7 \\
% 2018 Jan 27       & 162.9              & VLA       & 3     &  96.2    & 13.5 \\
2017 Sep 08       & 22.36    & VLA       & 3     & 22.5     & 3.4 \\
2017 Sep 10       & 24.26    & VLA       & 3     & 25.6     & 2.9 \\
2017 Sep 17       & 31.33    & VLA       & 3     & 34.0     & 3.6 \\
2017 Oct 02       & 46.25    & VLA       & 3     & 44       & 4 \\
2017 Oct 10       & 54.29    & VLA       & 3     & 48       & 6 \\
2017 Oct 13       & 57.19    & VLA       & 3     & 61       & 9 \\
2017 Nov 18       & 93.13    & VLA       & 3     & 70       & 6 \\
\hline
2017 Sep 05  & 18.66              & ATCA      & 7.25    & 20.0   & 4.8 \\
2017 Nov 01  & 75.49              & ATCA      & 7.25    & 35.9    & 4.3 \\
2017 Nov 17  & 92.4               & ATCA      & 7.25    & 31.7    & 5.6 \\
2017 Dec 02  & 107.36             & ATCA      & 7.25    & 53.2    & 4.5 \\
2017 Dec 20  & 125.3              & ATCA      & 7.25    & 58.2    & 5.0 \\
2018 Jan 13  & 149.26             & ATCA      & 7.25    & 60.6    & 4.3 \\
2018 Feb 01  & 181.64             & ATCA      & 7.25    & 57.9    & 6.9 \\
\hline
2017 Oct 21  & 65.14        &  GMRT    & 0.61     & 117   & 42 \\
\hline
\multicolumn{6}{p{3.5in}}{Notes: 1. Compilation from: \cite{2017Sci...358.1579H,2018Natur.554..207M,2018ApJ...858L..15D}. 2. The ATCA flux densities of GW170817 have a correction factor of 1.25 applied (see \S\ref{subsec:spectral} for details).}\\ 
\multicolumn{6}{l}{\dag~Mean epoch, days post-merger.}\\
\end{tabular}
\end{table}

\acknowledgements
%We thank P.~Chang for helpful discussions.  
Acknowledgements: The National Radio Astronomy Observatory is a facility of the National Science Foundation operated under cooperative agreement by Associated Universities, Inc. The Australia Telescope is funded by the Commonwealth of Australia for operation as a National Facility managed by CSIRO. 
We thank the staff of the GMRT that made these observations possible. GMRT is run by the National Centre for Radio Astrophysics of the Tata Institute of Fundamental Research.
The MeerKAT telescope is operated by the South African Radio Astronomy Observatory, (SARAO), which is a facility of the National Research Foundation (NRF), an agency of the Department of Science and Technology. The authors thank the NRAO staff, especially Mark Claussen and Amy Mioduszewski, for scheduling the VLA observations. K.P.M. would like to thank Kenta Hotokezaka and Michael Zhang for help with {\tt emcee}.  We thank Varun Bhalerao for help with some of the uGMRT observation 
AC acknowledges support from the NSF CAREER award \#1455090. DD is supported by an Australian Government Research Training Program Scholarship. K.P.M. is currently a Jansky Fellow of the National Radio Astronomy Observatory. TM acknowledges the support of the Australian Research Council through grant FT150100099. 
% Part of this research was conducted by the Australian Research Council Centre of Excellence for All-sky Astrophysics (CAASTRO), through project number CE110001020.
Part of this work was supported by the GROWTH (Global Relay of Observatories Watching Transients Happen) project funded by the National Science Foundation under PIRE Grant No 1545949. DK was additionally supported by by NSF grant AST-1412421. P.C. acknowledges support from the Department of Science and Technology via SwaranaJayanti Fellowship award (file no.DST/SJF/PSA-01/2014-15). 
The research of O.S. is supported by the South African Research Chairs Initiative of the Department of Science and Technology and NRF.

\facility{VLA, ATCA, uGMRT, MeerKAT}
\software{Matplotlib~\citep{hunter2007},~NumPy~\citep{oliphant2006}, 
% Astropy \citep{2018arXiv180102634T}, 
MIRIAD~\citep{1995ASPC...77..433S},~DIFMAP~\citep{1997ASPC..125...77S},~CASA \citep{2007ASPC..376..127M}},~{\tt emcee}~\citep{foreman-mackey2013},~{\tt corner}~\citep{foreman-mackey2016}
% * <ddob1600@uni.sydney.edu.au> 2018-07-20T06:40:21.802Z:
% 
% corner.py should be acknowledged separately from emcee
% 
% ^.

%\bibliography{bibliography}

\end{document}